\documentclass[9pt]{article}
\usepackage{spconf,amsmath,graphicx,algorithm}
\usepackage{amssymb}
\usepackage{stmaryrd}
\usepackage[mathscr]{eucal}
\usepackage{subfigure}
\usepackage[noend]{algorithmic}

\title{TWOFOLD VIDEO HASHING WITH AUTOMATIC SYNCHRONIZATION}
\name{Mu Li, Vishal Monga}
\address{Department of Electrical Engineering, The Pennsylvania State University, USA.}
\begin{document}
%
\maketitle
\begin{abstract}
Video hashing finds a wide array of applications in content authentication, robust retrieval and anti-piracy search. While much of the existing research has focused on extracting robust and secure content descriptors, a significant open challenge still remains: Most existing video hashing methods are fallible to temporal desynchronization. That is, when the query video results by deleting or inserting some frames from the reference video, most existing methods assume the positions of the deleted (or inserted) frames are either perfectly known or reliably estimated. This assumption may be okay under typical transcoding and frame-rate changes but is highly inappropriate in adversarial scenarios such as anti-piracy video search. For example, an illegal uploader will try to bypass the `piracy check' mechanism of YouTube/Dailymotion etc by performing a cleverly designed non-uniform resampling of the video. We present a new solution based on dynamic time warping (DTW), which can implement automatic synchronization and can be used together with existing video hashing methods. The second contribution of this paper is to propose a new robust feature extraction method called flow hashing (FH), based on frame averaging and optical flow descriptors. Finally, a fusion mechanism called distance boosting is proposed to combine the information extracted by DTW and FH. Experiments on real video collections show that such a hash extraction and comparison enables unprecedented robustness under both spatial and temporal attacks.

\end{abstract}
%
%
\section{INTRODUCTION}
\label{sec:intro}

Video hashing is a dimensionality reduction technique which transforms a raw video to a compact vector that can facilitate content-based retrieval. Other applications include video authentication, anti-piracy search and augmented reality. Robustness against content-preserving distortions is a central requirement of video hashing, and security applications often also require cryptographic key based randomization.

\noindent \textbf{Existing literature and motivations of this paper}:  The existing video hashing techniques can be roughly classified into two types. The first type, lower order information methods,  extract hash vectors directly from the video frames. The typical methods of this type include Radial hASHing (RASH, \cite{Roover05TSP}), Discrete Cosine Transform (DCT, \cite{Coskun06DCT}), Centroids of Gradient Orientations (CGO, \cite{Lee08CGO}) and Temporally Informative Representative Images (TIRI, \cite{Esmaeili11TIFS}). These methods either extract some geometric information such as RASH (which samples each frame using a set of lines centered at the frame's midpoint) and CGO (which computes local gradients), or calculate some transform coefficients (usually discrete cosine transform due to its energy compaction property) like DCT and TIRI. Note that TIRI proposed frame averaging during hash extraction and showed this operation is very robust to temporal distortions. Although lower order information gained initial success, researchers discovered that higher order information can achieve even better performance. Methods of this type extract hash vectors not directly from video frames, but from correlations between nearby frames. One representative method in this type is HOOF \cite{Chaudhry09CVPR, Ren13TIFS}, where Histogram of Orientations of Optical Flow is used to extract hash vectors. Other advances include the use of multiple hash vectors to generate binary hash bits using spectral hashing \cite{Song13TM, Lv13TIFS}. A practical challenge with \cite{Song13TM, Lv13TIFS} is that as sufficient number of new videos are added to the database, retraining is needed and all hash vectors must be regenerated. Our goal is instead to develop fusion techniques such that model retraining does not influence existing hashes in the database. The central challenge we seek to overcome is the open problem of temporal desynchronization in video hashing. There have indeed been notable attempts in this direction, namely in  \cite{cano2005review,harmanci2005temporal} where frame based image hashes can be used to synchronize audio or video. But these techniques invariably require complicated combinatorial optimization problem and are hence quite expensive. Further, the strategy to normalize the query video to the same length with reference video even after finding correspondence does not seem unique.

\noindent \textbf{Contributions of this paper}: This paper develops a new video hashing paradigm called: Twofold hashing. First, a preprocessing method based on DTW is implemented, which can automatically detect the positions of deleted /inserted frames quickly and reliably synchronize the distorted query video to the same length with the original reference video. Note that the proposed preprocessing is universal in the sense that it can be used together with virtually any robust feature extractor in existing video hashing methods. Next, we propose a new robust feature extractor called flow hashing (FH), which tries to blend the frame averaging operation in \cite{Esmaeili11TIFS} with HOOF in \cite{Ren13TIFS}. Finally, we propose a fusion technique called distance boosting which aims to combine (fuse) the measure given by DTW-based preprocessing and Euclidean distance between FH hashes. Experiments confirm that the detection performance in terms of Receiver Operating Characteristics (ROCs), is significantly improved when compared against state of the art video hashing techniques.

\section{ALGORITHM FORMULATION}
\label{sec:formu}

\begin{figure}
  \centering
  \includegraphics[width=9.5cm]{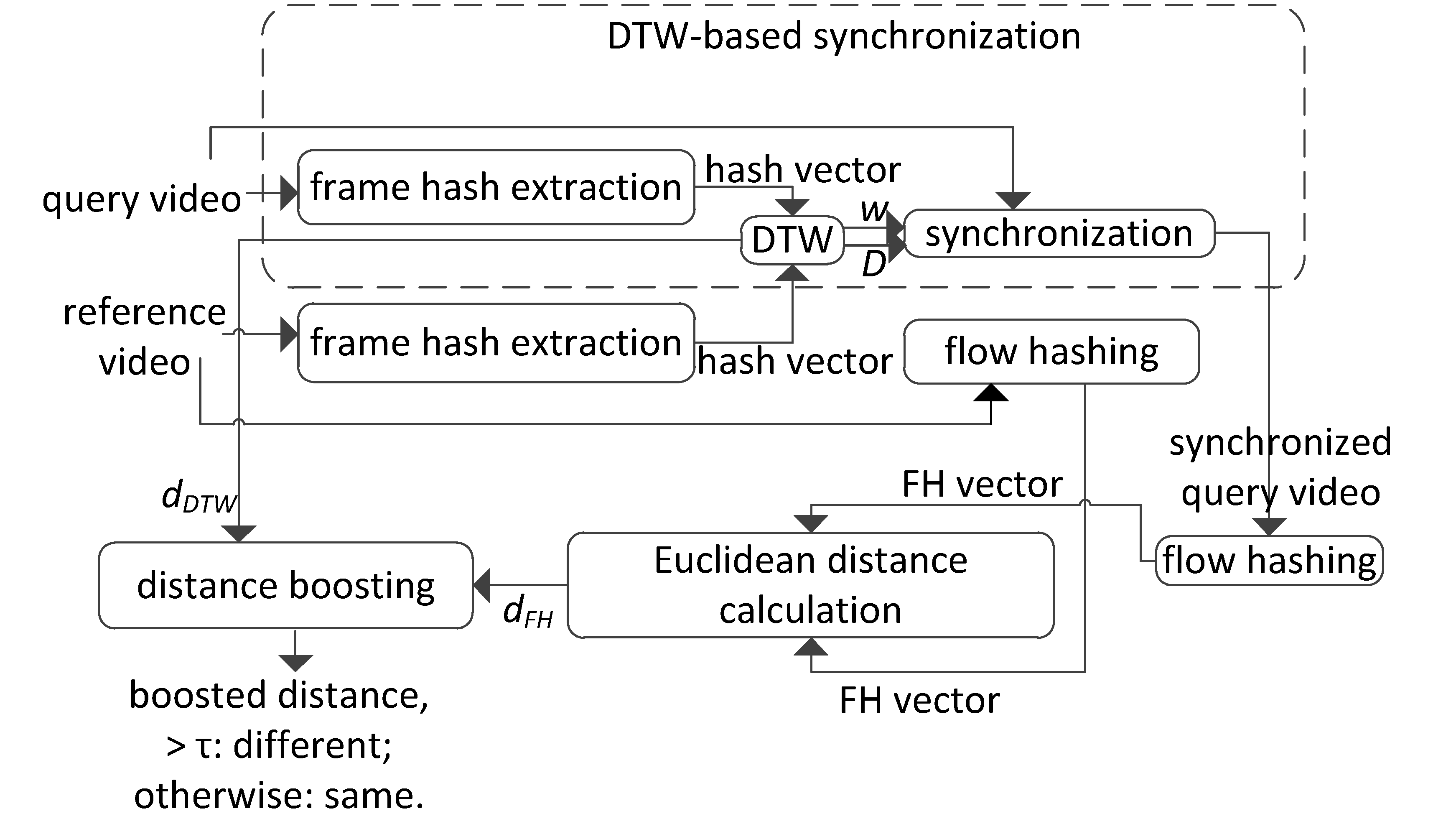}\\
  \caption{Proposed twofold video hashing system.}\label{fig:stru}
\end{figure}

\subsection{DTW-based video synchronization}
\label{sec:formu:dtw}

The central idea of the proposed synchronization is: Extract hash vectors from each frame, model the resulting hash vector as time series, and apply dynamic time warping to synchronize query time series to reference time series. Dynamic time warping \cite{vintsyuk1968DTW} is based on dynamic programming principle \cite{bellman1953DP}, whose central idea is: Given the starting point, the problem of finding the optimal path to an end point is equivalent to first go to an optimal middle point and then find the optimal path starting from that middle point.

\begin{figure}
\begin{minipage}[c]{.45\linewidth}
\subfigure[]{
\includegraphics[width=4.3cm,height=3cm]{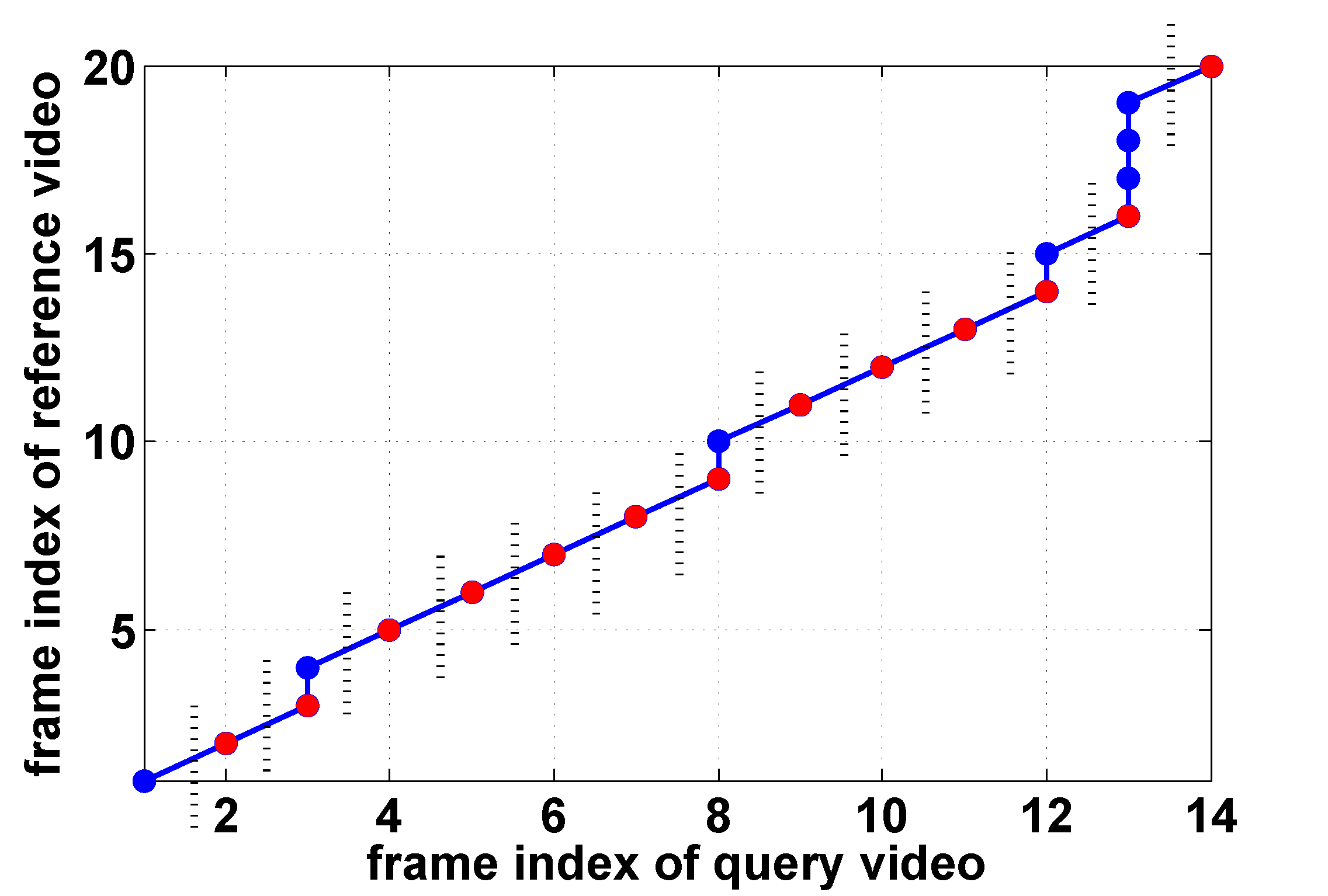}
}
\caption{An example warping path.}
\label{fig:path}
\end{minipage}
\quad
\begin{minipage}[c]{.45\linewidth}
\subfigure[]{
  \includegraphics[height=2.0cm]{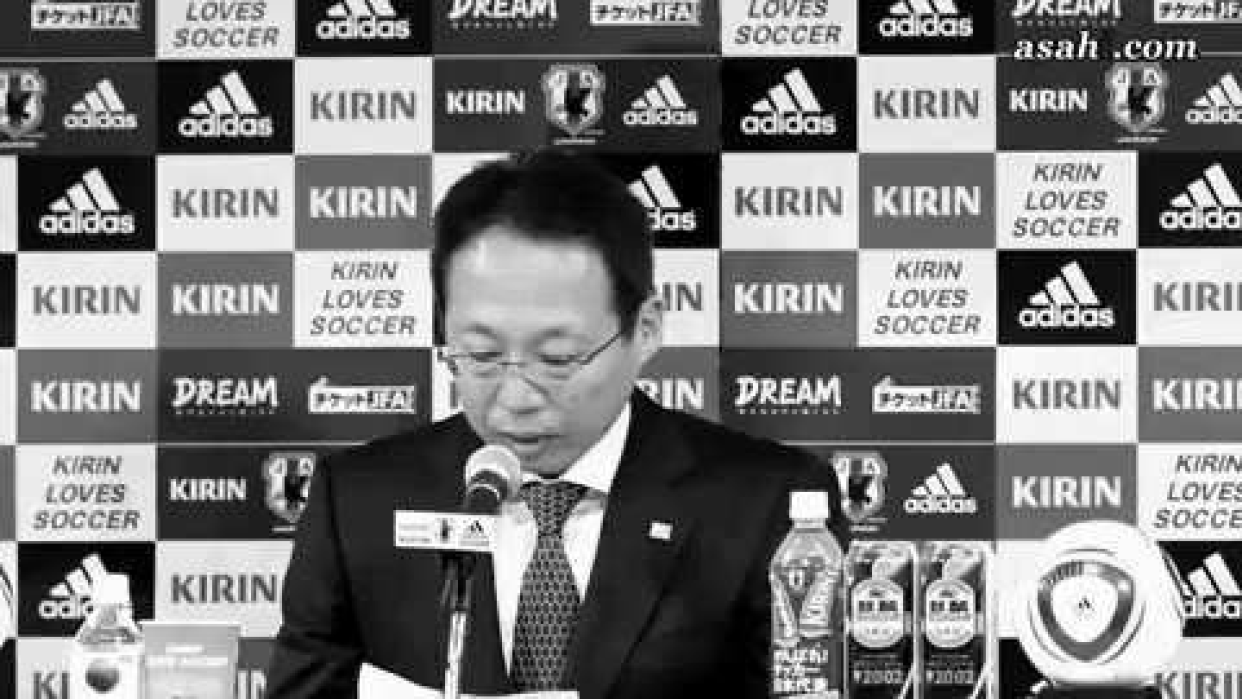}\label{fig:sam:s1}
}
\subfigure[]{
\includegraphics[height=2.0cm]{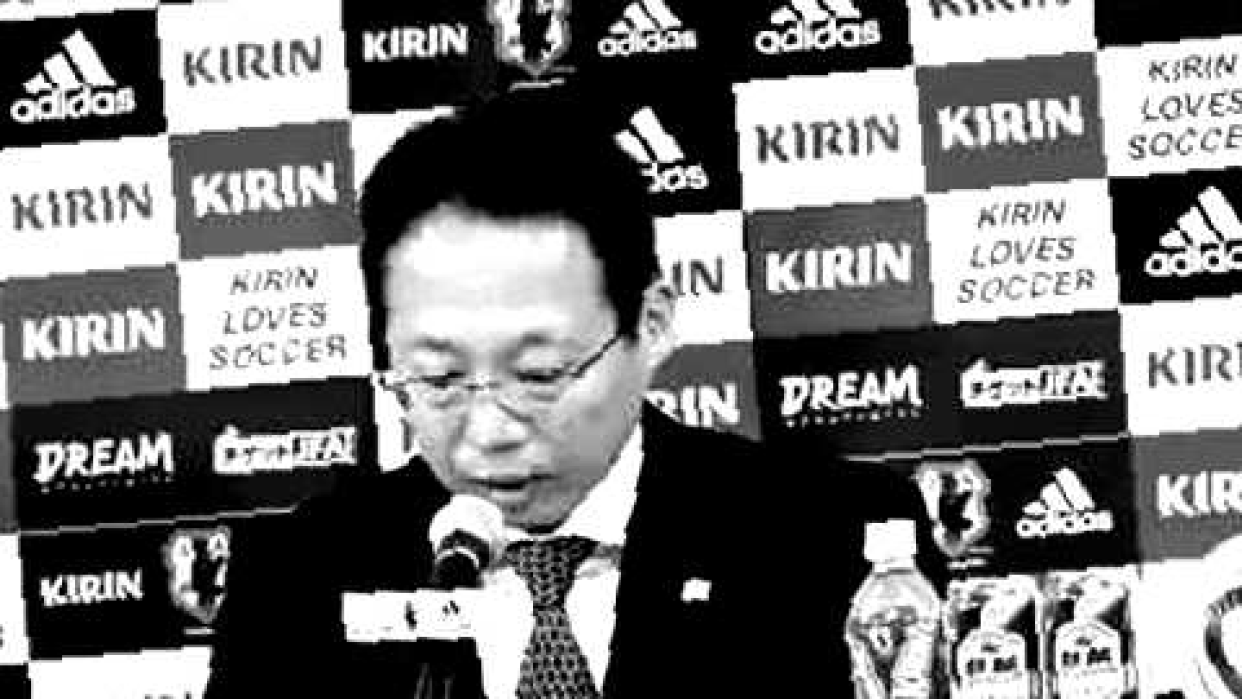}\label{fig:sam:s2}
}
\caption{A typical frame: (a) Original; (b) Attacked/Distorted.}
\label{fig:sam}
\end{minipage}
\end{figure}

Specifically, we model two frame-based hash vectors $\mathbf{h}_{r}^{f}$ (from reference video $\boldsymbol{\mathscr{V}}_r$) and $\mathbf{h}_{q}^{f}$ (from query video $\boldsymbol{\mathscr{V}}_q$) as time series, apply DTW to compute an optimal warping path $w$ and a distance $d_{\mathrm{DTW}}$ simultaneously. Here $w$ represents an correspondence between $\mathbf{h}_{r}^{f}$ and $\mathbf{h}_{q}^{f}$. A warping path is denoted by
\begin{equation}
  w=\{(i_k,j_k)\}_{k=1}^p,
\end{equation}
which represents $i_k$'th frame in the query video corresponds to $j_k$'th frame in the reference video. An example warping path is shown in Fig. \ref{fig:path}, where the red points (point whose $x-$coordinate and $y-$coordinate are increased by 1 simultaneously comparing with the previous point on warping path) are of particular interest since they represent the beginning points of matching intervals that are separated by black dashed lines. DTW solves the following problem:
\begin{equation}\label{equ:d}
  d_{\mathrm{DTW}}\doteq\min_w(\sum_{k=1}^p D(i_{k},j_{k})),
\end{equation}
where $D(\cdot,\cdot)$ is the basic metric (defined in Step \ref{dtw:step} of Algorithm \ref{alg:dtwsyn} in our scenario). Solving (\ref{equ:d}) is unfortunately combinatorially explosive; the standard DTW uses some constraints such as monotonicity and continuity to reduce size of search space. As a result, (\ref{equ:dtw}) is used to approximate (\ref{equ:d}):
\begin{equation}\label{equ:dtw}
\begin{aligned}
    &\gamma(i,j)=\underbrace{D(i,j)+\min}_{\substack{\text{find the optimal present step}\\\text{ }\\ \text{ }}}(\\ &\underbrace{[\gamma(i-1,j),\gamma(i-1,j-1),\gamma(i,j-1)]}_{\substack{\text{warping path is optimal until the previous step}\\\text{} \\\text{ }}}).
\end{aligned}
\end{equation}
The warping path $w$ is computed by the index of the minimizers in (\ref{equ:dtw}) backwardly, and $d_{\mathrm{DTW}}\approx\gamma(\frac{\mathrm{length}(\mathbf{h}_{r}^{f})}{2},\frac{\mathrm{length}(\mathbf{h}_{q}^{f})}{2}).$
Note that $d_{\mathrm{DTW}}$ is not a ``metric" strictly because it allows two different points to have zero distance and does not satisfy triangle inequality. But $d_{\mathrm{DTW}}$ can measure how close two vectors are in our scenario. DTW has long been known as one of the most powerful ways of comparing time-series of different lengths. Further, recent algorithmic advances ensure that DTWs can be computed very fast \cite{salvador2007toward}.
We leverage these merits into developing a video synchronization method as stated in Algorithm \ref{alg:dtwsyn}.

\begin{algorithm}
\caption{DTW-based video synchronization}\label{alg:dtwsyn}
\begin{algorithmic}[1]
\STATE [\textbf{Frame hash extraction}]: Apply $2-$D DCT to frames of query video $\boldsymbol{\mathscr{V}}_q$, extract the first horizontal and vertical coefficients (adjacent to DC) to form $\mathbf{h}_{q}^{f}$. The corresponding $\mathbf{h}_{r}^{f}$ for reference $\boldsymbol{\mathscr{V}}_r$ was computed offline.
\STATE \COMMENT {\%\textbf{comment: Given $\mathbf{h}_{r}^{f}$ and $\mathbf{h}_{q}^{f}$, Step \ref{alg:syn:be} -Step \ref{alg:syn:end} apply DTW to compute $w$, $D$ and $d_{\mathrm{DTW}}$.}\%}
\STATE \COMMENT {\%\textbf{comment: Compute $\ell_2$ distance between every two frames from $\boldsymbol{\mathscr{V}}_r$ and $\boldsymbol{\mathscr{V}}_q$:}\%}\label{alg:syn:be}
\FOR {each $n \in \{1,2,...,\frac{\mathrm{length}(\mathbf{h}_{q}^{f})}{2}\} $}
\FOR {each $m \in \{1,2,...,\frac{\mathrm{length}(\mathbf{h}_{r}^{f})}{2}\} $}
     \STATE $D(n,m)=\|\mathbf{h}_{q}^{f}(1+(n-1)\cdot 2:2\cdot n)-\mathbf{h}_{r}^{f}(1+(m-1)\cdot 2:2\cdot m)\|_2$. \label{dtw:step}	
 \ENDFOR
\ENDFOR
\STATE [\textbf{Standard DTW}]: Given $D$, apply Equ. (\ref{equ:dtw}) to get optimal warping path $w$ and distance $d_{\mathrm{DTW}}$.\label{alg:syn:end}
\STATE \COMMENT {\%\textbf{comment: Given $w$ and $D$, Step \ref{alg:syn:Beg} -Step \ref{alg:syn:End} synchronize $\boldsymbol{\mathscr{V}}_q$ to the same length as $\boldsymbol{\mathscr{V}}_r$, the synchronized video is saved in $\boldsymbol{\mathscr{V}}_q^{syn}$}.\%}
\STATE [\textbf{Extract beginning points}]: Given $w$, extract the coordinates of the $T$ matching intervals' beginning points (red points in Fig. \ref{fig:path}), $\{x_i^1,y_i^1\}_{i=1}^T$, where $x_i^1$ is a frame index in $\boldsymbol{\mathscr{V}}_q$, $y_i^1$ is a frame index in $\boldsymbol{\mathscr{V}}_r$.\label{alg:syn:Beg}
\STATE [\textbf{Extract matching intervals}]: From beginning points, extract coordinates of other points in the same matching interval (blues points between the same two dashed lines with each red point in Fig. \ref{fig:path}), $\{\{x_i^j,y_i^j\}_{j=1}^{B_i}\}_{i=1}^T$, $B_i$ is the number of points in $i$'th interval.
\FOR {each $i \in \{1,2,...,T\} $}
     \STATE $Mat(i,:)=[x_i^p,y_i^p]$ where $D(x_i^p,y_i^p)$ is minimal among $\{D(x_i^j,y_i^j)\}_{j=1}^{B_i}$.\COMMENT {\%\textbf{comment: $Mat()$ stores the coordinates of matching points; from each interval, choose matching point to be $(x_i^p,y_i^p)$ such that $D(x_i^p,y_i^p)$ is smallest in the same interval}.\%}	
 \ENDFOR
\FOR {each $i \in \{1,2,...,T\} $}
     \STATE $\boldsymbol{\mathscr{V}}_q^{syn}(:,:,Mat(i,2))=\boldsymbol{\mathscr{V}}_q(:,:,Mat(i,1))$.
 \ENDFOR
\STATE [\textbf{Interpolation}]: Apply interpolation if there are missing frames in the middle of $\boldsymbol{\mathscr{V}}_q^{syn}$.
\STATE [\textbf{Extrapolation}]: Apply extrapolation if there are missing frames in the beginning or end of $\boldsymbol{\mathscr{V}}_q^{syn}$.\label{alg:syn:End}
\end{algorithmic}
\end{algorithm}
\begin{algorithm}
\caption{flow hashing}\label{alg:fh}
\begin{algorithmic}[1]
\STATE [\textbf{Frame averaging}]: Given the input video $\boldsymbol{\mathscr{V}}_{in}$,
\FOR {each $j \in \{1,2,...,\frac{length(\boldsymbol{\mathscr{V}}_{in})}{J}\} $}
     \STATE $TIRI(:,:,j)=\frac{1}{J}\sum_{k=1+(j-1)\cdot J}^{j\cdot J}{\boldsymbol{\mathscr{V}}}_{in}(:,:,k)$.
     \COMMENT {\%comment: Frame averaging; it can also be done in overlapping segments.}
 \ENDFOR
 \STATE [\textbf{Compute optical flow}]: Calculate optical flow between every two successive $TIRI$, and the histogram of orientations of the optical flow, counted by motion amplitude. Concatenate all histograms to form vector $\mathbf{h}_{in}^{o}$.
 \STATE [\textbf{Normalization}]: $\mathbf{h}_{in}^{o}=\frac{\mathbf{h}_{in}^{o}}{\parallel\mathbf{h}_{in}^{o}\parallel_2}$.
\end{algorithmic}
\end{algorithm}

\vspace{-4mm}
\subsection{Robust representation: Flow hashing (FH)}
\label{sec:formu:fh}
The central idea of flow hashing is to blend the frame averaging operation in \cite{Esmaeili11TIFS} and HOOF feature in \cite{Chaudhry09CVPR}. Frame averaging operation is shown to be robust to temporal attacks and can also help to reduce hash length \cite{Esmaeili11TIFS}; HOOF is used because the pixel motions, as a higher order information, is one of the most definitive features of a video. Specifically, an optical flow is a vector field, which is a map $X$ from a manifold $M$ to its tangent bundle $TM$,
\begin{equation}
\begin{aligned}
    X: \underbrace{M}_{\substack{\text{a manifold}\\\text{ }\\ \text{ }}} \rightarrow\underbrace{TM}_{\substack{\text{tangent bundle }\\\text{of the manifold} \\\text{ }}}.
\end{aligned}
\end{equation}
In our algorithm, $M$ is $\mathbb{R}^2$ and we use \cite{Bruhn05IJCV} to compute optical flow $X$; and the resulted optical flow is encoded through histogram of orientations counted by motion amplitude. The formal steps of FH are described in Algorithm \ref{alg:fh}.
\vspace{-2mm}
\subsection{Distance boosting}
\label{sec:formu:db}

We propose distance boosting, which tries to fuse different distances through linear combination as AdaBoost \cite{freund1997decision} does for classifiers. Both frame-based hashes $\mathbf{h}^{f}$ and flow hashes $\mathbf{h}^{o}$ are utilized so that the overall detection performance can be improved. In the training phase:
\begin{equation}\label{equ:md}
  d_{\mathrm{DTW}}(\boldsymbol{\mathscr{V}}_r,\boldsymbol{\mathscr{V}}_q)\doteq\gamma(\frac{\mathrm{length}(\mathbf{h}_{r}^{f})}{2},\frac{\mathrm{length}(\mathbf{h}_{q}^{f})}{2}),
\end{equation}
\begin{equation}\label{equ:mf}
  d_{\mathrm{FH}}(\boldsymbol{\mathscr{V}}_r,\boldsymbol{\mathscr{V}}_q)\doteq\parallel\mathbf{h}_{r}^{o}-\mathbf{h}_{q}^{o}\parallel_2,
\end{equation}
\begin{equation}\label{equ:mb}
  d_{\mathrm{boost}}(\boldsymbol{\mathscr{V}}_r,\boldsymbol{\mathscr{V}}_q)\doteq\alpha_1\cdot d_{\mathrm{DTW}}(\boldsymbol{\mathscr{V}}_r,\boldsymbol{\mathscr{V}}_q)+\alpha_2\cdot d_{\mathrm{FH}}(\boldsymbol{\mathscr{V}}_r,\boldsymbol{\mathscr{V}}_q),
\end{equation}
we try to make $d_{\mathrm{boost}}$ between visually same videos smaller than that between visually different videos, as much as possible. Slack variable technique \cite{bishop2006pattern} used in $1-$norm soft margin SVM is used here to realize this goal, i.e., we solve
\begin{equation}\label{equ:train}
  \begin{aligned}
& \underset{\alpha_1,\alpha_2}{\min}
& & \sum_{(i,j,k)\in \boldsymbol{\mathscr{I}}}[\left(\alpha_1\cdot d_{\mathrm{DTW}}(\mathbf{h}_i^f,\mathbf{h}_j^f)+\alpha_2\cdot d_{\mathrm{FH}}(\mathbf{h}_i^o,\mathbf{h}_j^o)\right)\\
& & &+1-\left(\alpha_1\cdot d_{\mathrm{DTW}}(\mathbf{h}_i^f,\mathbf{h}_k^f)+\alpha_2\cdot d_{\mathrm{FH}}(\mathbf{h}_i^o,\mathbf{h}_k^o)\right)]_{+} \\
& \text{s.t.}
& & \alpha_1\geq0, \alpha_2\geq0.
\end{aligned}
\end{equation}
\begin{equation}
\begin{aligned}
 & \boldsymbol{\mathscr{I}}\doteq\{(i,j,k): \boldsymbol{\mathscr{V}}_i \text{ and } \boldsymbol{\mathscr{V}}_j \text{ are visually same videos},\\
  &\boldsymbol{\mathscr{V}}_i \text{ and } \boldsymbol{\mathscr{V}}_k \text{ are visually different videos.}\}
  \end{aligned}.
\end{equation}
Note that since (\ref{equ:train}) is convex (essentially equivalent to a linear program), fast numerical techniques yield the optimal solution. Note that distance boosting is inspired by the distance combination technique proposed in \cite{Jang13ICCSPA}, which is based on distance metric learning. The differences from \cite{Jang13ICCSPA} are: 1) We use a different objective function which, unlike \cite{Jang13ICCSPA}, does not need cross-validation; 2) in distance boosting, $d_{\mathrm{DTW}}$ is fused, which is not a metric. Further in comparison with spectral hashing techniques \cite{Song13TM, Lv13TIFS}, distance boosting only needs to update $\alpha_1$ and $\alpha_2$ rather than having to regenerate all hash vectors under retraining.

\section{EXPERIMENTS AND ANALYSIS}
\label{sec:exp}


Detection performance is measured by ROCs of a binary hypothesis test problem where $H_1$ assumes the query video $\boldsymbol{\mathscr{V}}_q$ is a distorted version of the reference video $\boldsymbol{\mathscr{V}}_r$; $H_0$ assumes $\boldsymbol{\mathscr{V}}_q$ and $\boldsymbol{\mathscr{V}}_r$ are visually different videos.
The error probabilities are defined as
\begin{equation}\label{equ:pm}
    P_M(\tau)=\mathrm{Pr}( d(\mathbf{H}(\boldsymbol{\mathscr{V}}),\mathbf{H}(A(\boldsymbol{\mathscr{V}}))) \geq\tau)
\end{equation}
\begin{equation}\label{equ:pa}
   P_{FA}(\tau)=\mathrm{Pr}( d(\mathbf{H}(\boldsymbol{\mathscr{V}}),\mathbf{H}(A(\boldsymbol{\mathscr{V}}'))) < \tau)
\end{equation}
where $A(\cdot)$ denotes content-preserving attacks. The distortions/attacks we test against are: $1$.) Spatial attack: Rotate $5$ deg, crop to $[\frac{3}{4}\text{Width}, \frac{3}{4}\text{Height}]$, intensity changes from $[0.2, 0.8]$ to $[0, 1]$; (The visual effect is shown in Fig. \ref{fig:sam}) $2$.) Temporal attack: $30\%$ of frames are dropped randomly and non-uniformly; $3$.) Spatio-temporal attack: Spatial attack plus Temporal attack as articulated above. We compare our algorithm with two widely cited methods in CGO \cite{Lee08CGO} and TIRI \cite{Esmaeili11TIFS}; the algorithm parameters are set so that FH, CGO and TIRI will produce hash vectors of roughly the same length ($64$, $80$, $72$, respectively). $700$ videos are downloaded from YouTube. In each simulation, $700$ matching video pairs and $700$ visually distinct video pairs are used. In FH with distance boosting, $350$ pairs are used in training and other $350$ pairs are used in testing. Each video is normalized to $64 \times64 \times2$ f$/$s.

\begin{figure}[h!]
\begin{minipage}[t]{.45\linewidth}
\subfigure[]{
  \includegraphics[width=4.3cm,height=3cm]{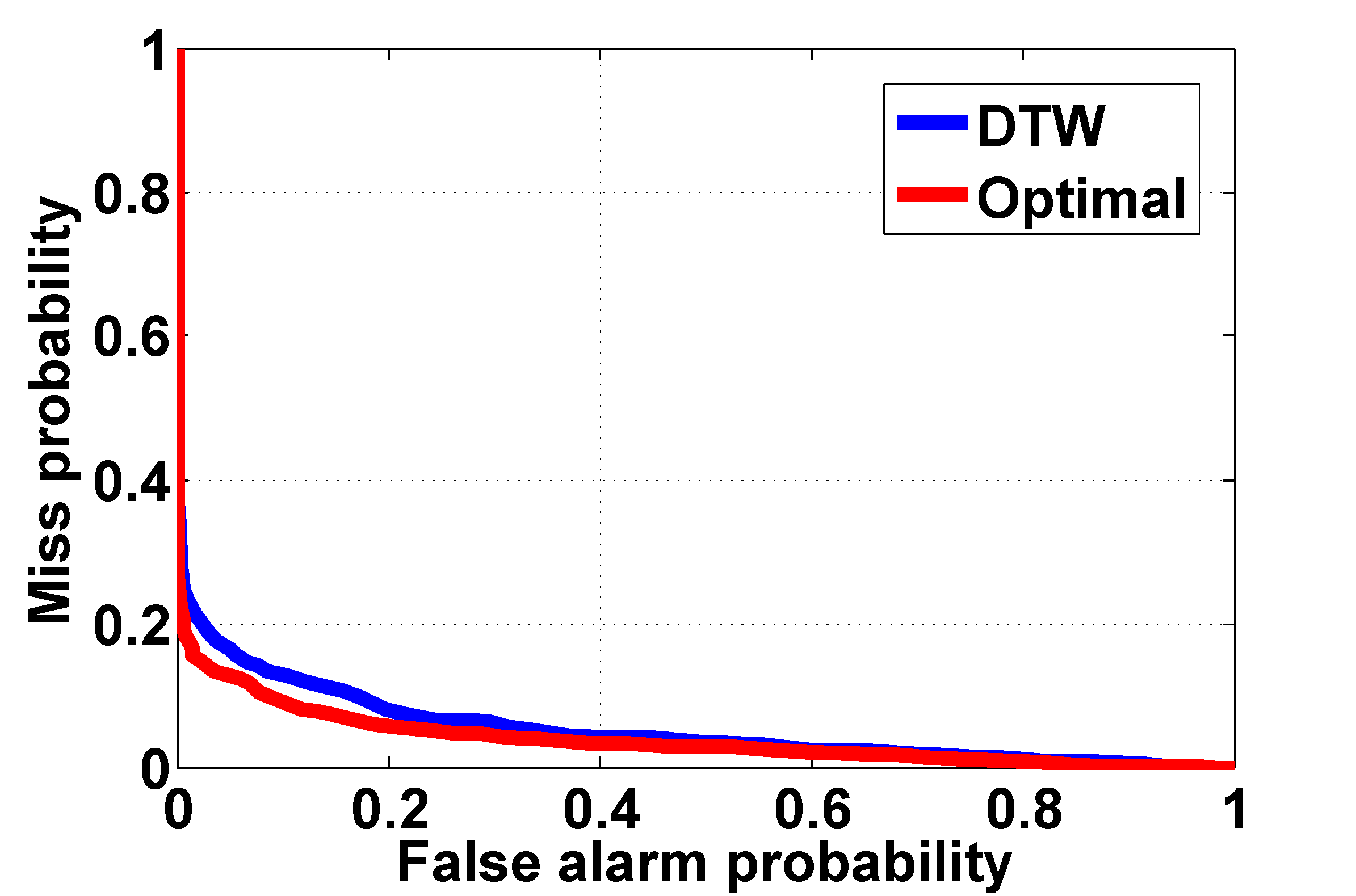}\label{fig:syn:s1}
}\\
\hspace{-5mm}
\subfigure[]{
\includegraphics[width=4.3cm,height=3cm]{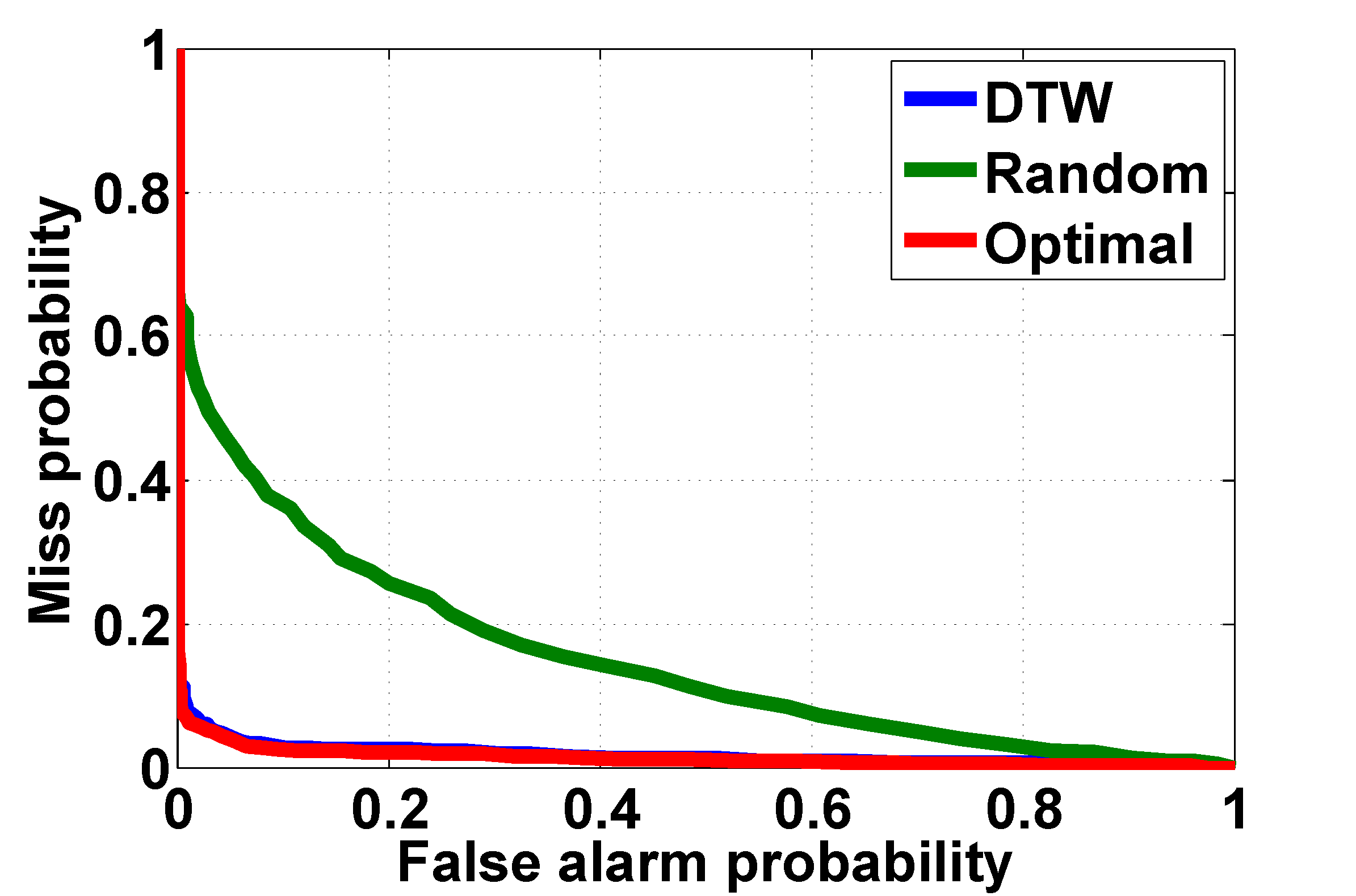}\label{fig:syn:s2}
}\\
\hspace{-5mm}
\subfigure[]{
\includegraphics[width=4.3cm,height=3cm]{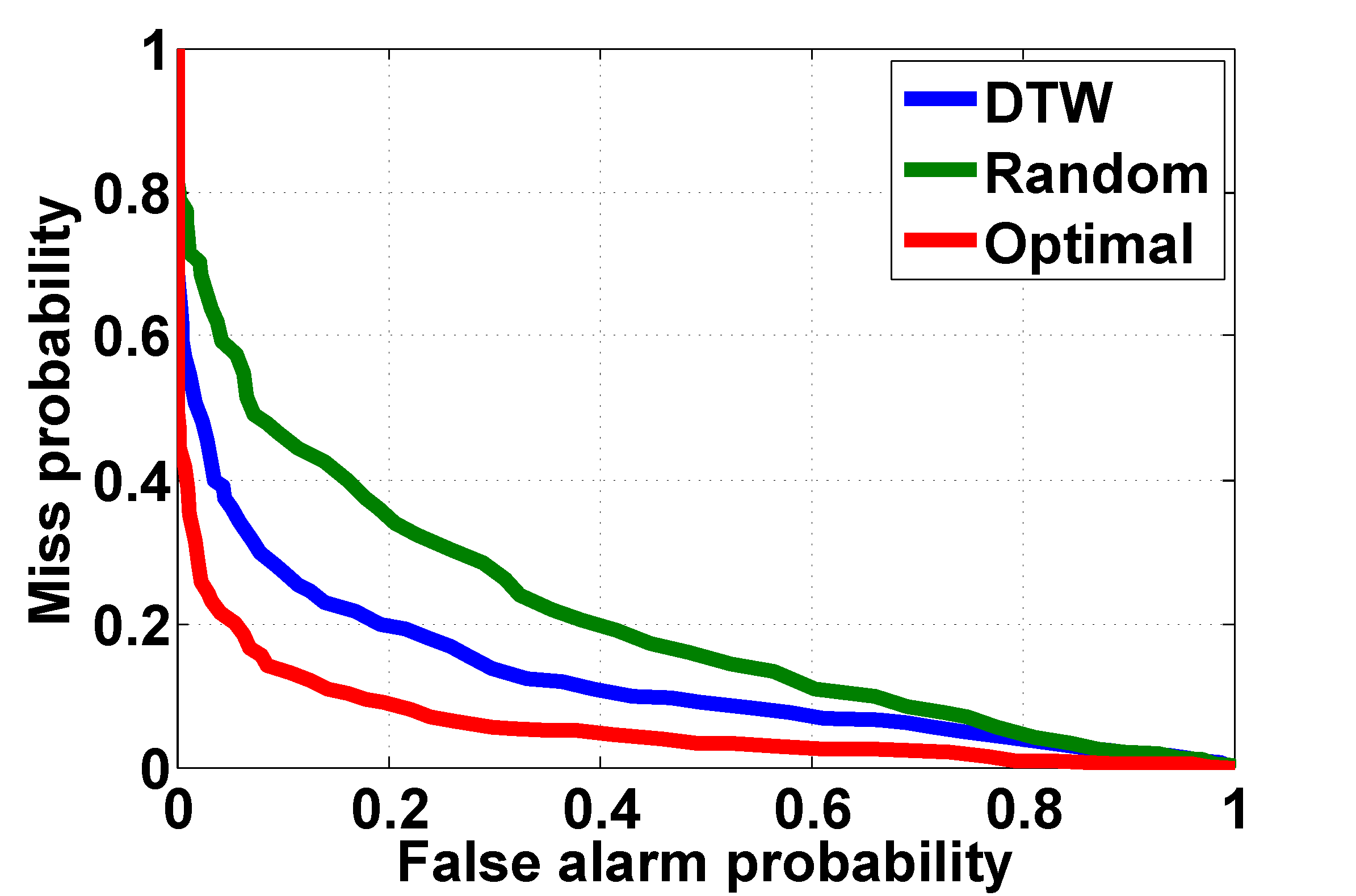}\label{fig:syn:s3}
}
\vspace{-2mm}
\caption{\small{Benefits of the proposed synchronization method to detection performance under: (a) Spatial attack; (b) Temporal attack; (c) Spatio-temporal attack.}}
\label{fig:syn}
\end{minipage}
\quad
\begin{minipage}[t]{.45\linewidth}
\subfigure[]{
  \includegraphics[width=4.3cm,height=3cm]{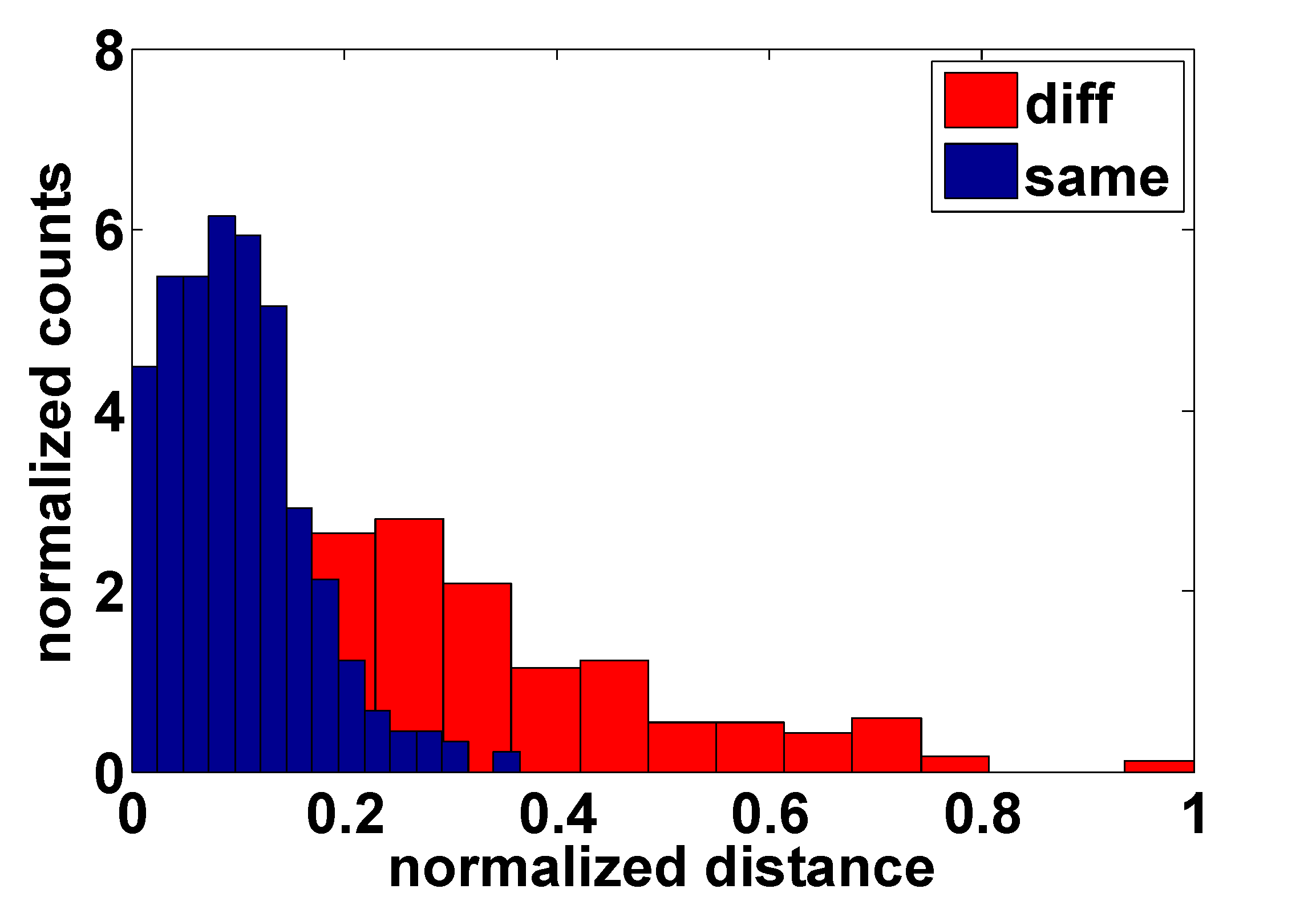}\label{fig:db:s1}
}
\subfigure[]{
\includegraphics[width=4.3cm,height=3cm]{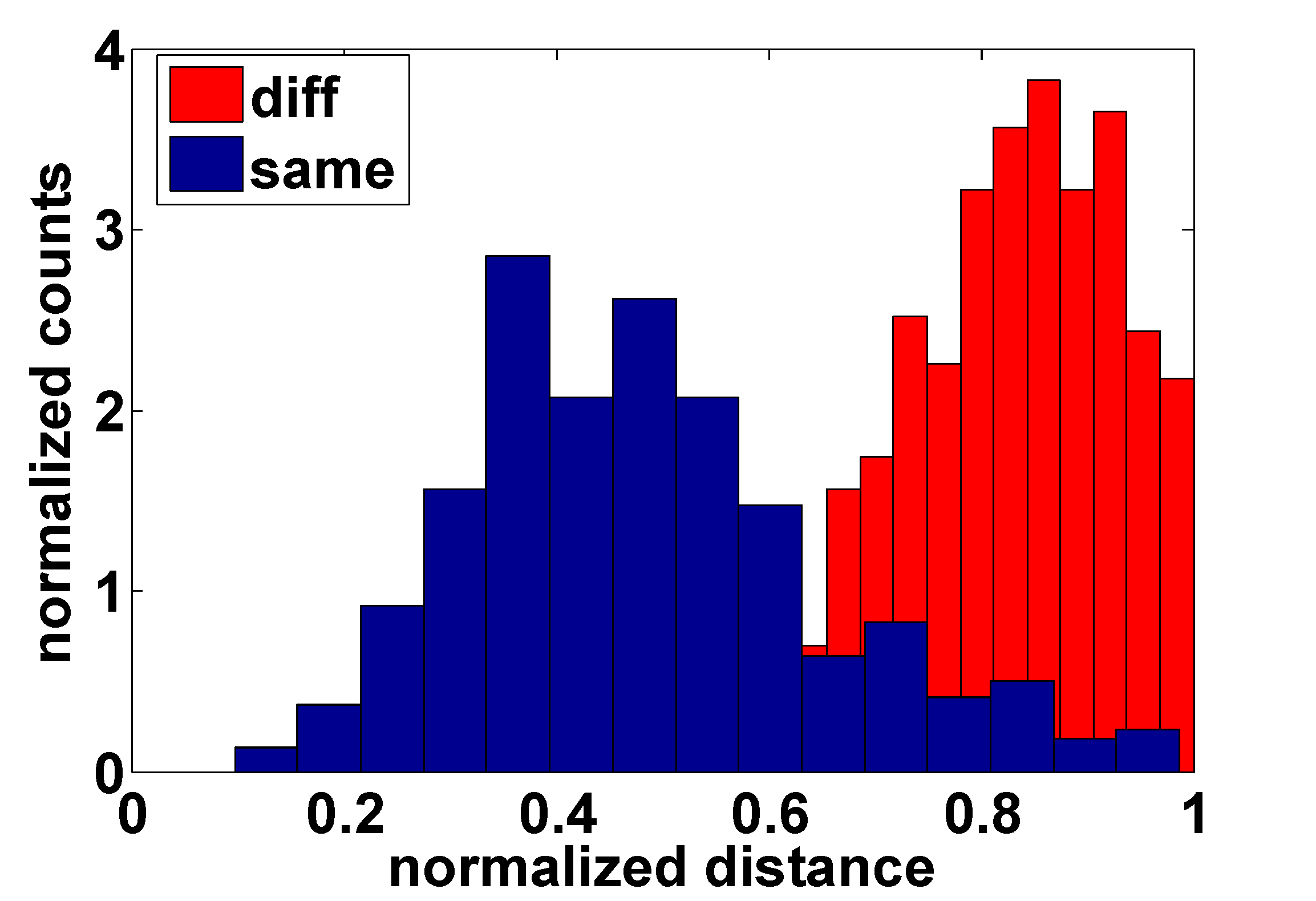}\label{fig:db:s2}
}
\subfigure[]{
\includegraphics[width=4.3cm,height=3cm]{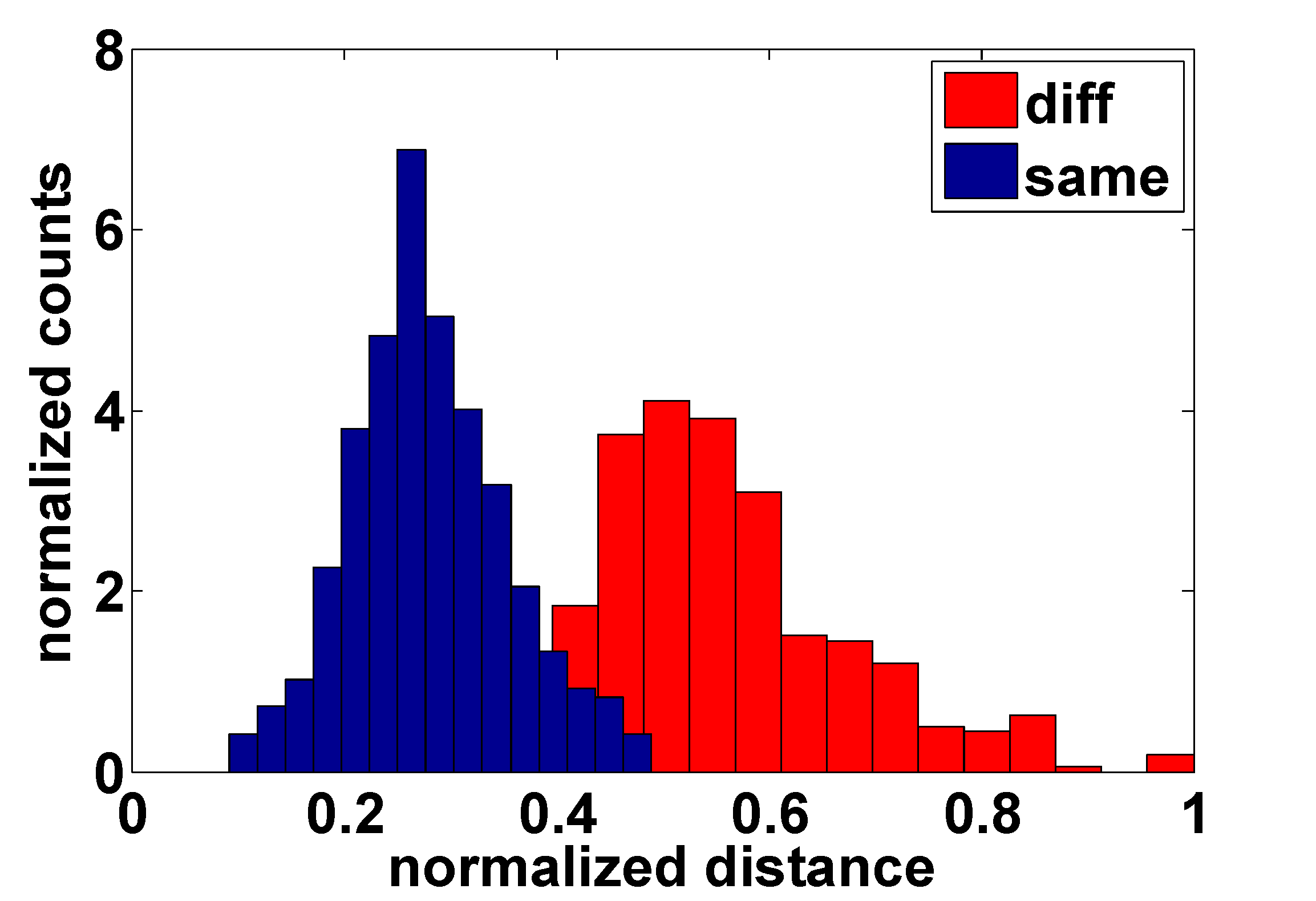}\label{fig:db:s3}
}
\vspace{-2mm}
\caption{\small{Histogram of distances between visually different videos (red) and similar videos (blue) (spatial attack): (a) DTW distances; (b) FH; (c) boosted distances.}}
\label{fig:db}
\end{minipage}
\begin{minipage}[t]{1\linewidth}
\centering
\subfigure{
\includegraphics[width=5cm]{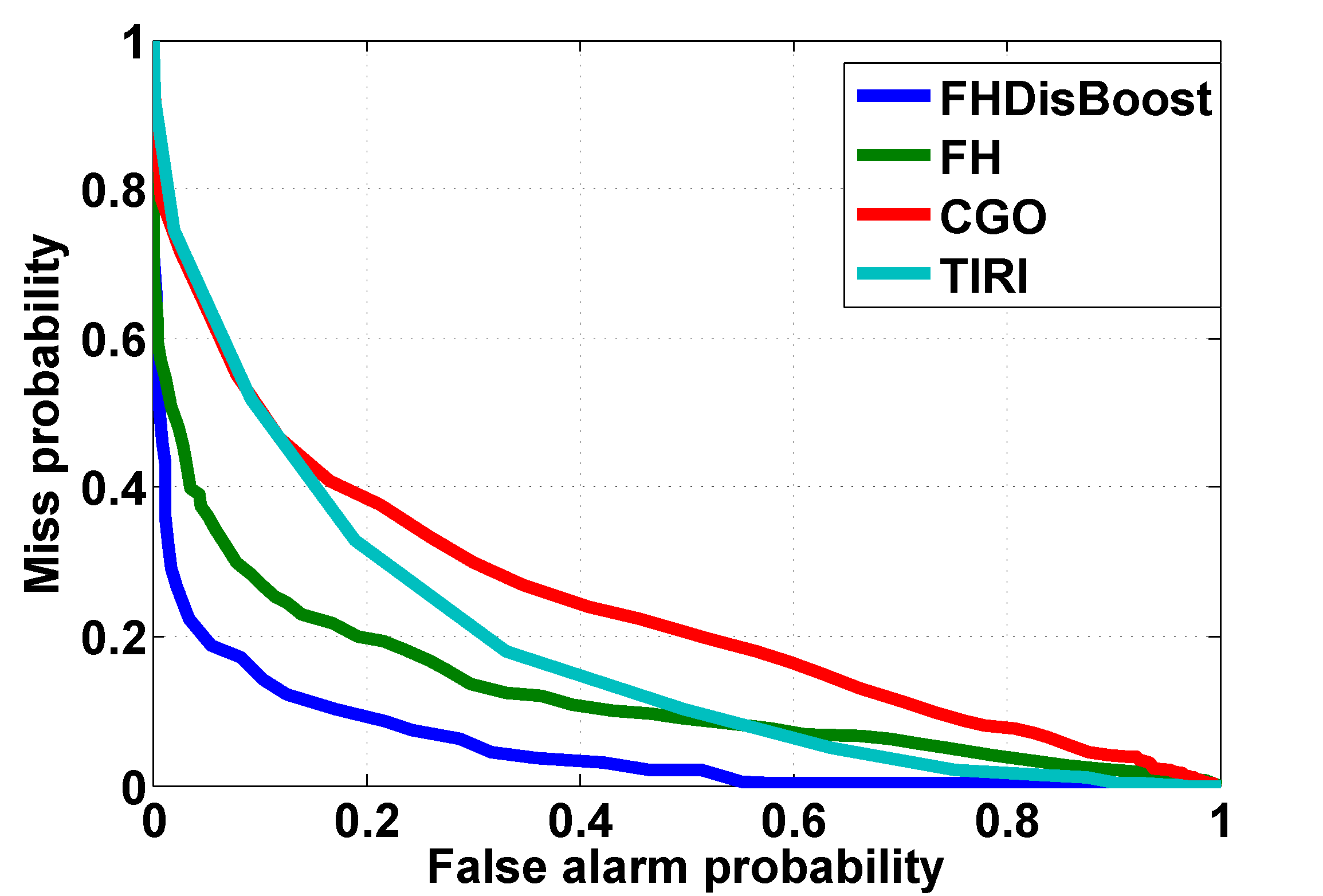}\label{fig:roc:s1}
}

\caption{\small{ROC comparisons against state of the art video hashing techniques: Spatio-temporal attack.}}
\label{fig:roc}
\end{minipage}
\end{figure}

\noindent \textbf{Benefits of automatic synchronization:} We choose flow hashing as $\mathbf{H}(\cdot)$, and Euclidean metric as $d(\cdot,\cdot)$, and plot the ROCs of three cases: $1$) DTW: Synchronized using the proposed method; $2$) optimal: Assume the frame deletion /insertion positions are perfectly known; $3$) random: Assume frames are deleted /inserted at random positions. Note that Case $2$ though unrealistic provides a bound to benchmark given methods since it leads to perfect synchronization; Case $3$ represents the current real-world scenario since YouTube for example doesn't know the deletion/insertion positions chosen by a potentially malicious uploader. Fig.\ \ref{fig:syn} reveals that the proposed method can significantly improve detection performance under temporal attacks in particular.

\noindent \textbf{Why distance boosting works:} We make different choices of $d(\cdot,\cdot)$ to explain why the proposed distance boosting can improve detection performance. We plot the histogram of normalized $d_{\mathrm{DTW}}$ in (\ref{equ:md}), $d_{\mathrm{FH}}$ in (\ref{equ:mf}) and $d_{\mathrm{boost}}$ in (\ref{equ:mb}), respectively, in Fig. \ref{fig:db}, from which we can see that if we only use either frame-based hashes $\mathbf{h}^{f}$ (from which $d_{\mathrm{DTW}}$ is calculated), or flow hashes $\mathbf{h}^{o}$ (from which $d_{\mathrm{FH}}$ is calculated), the histogram of distances between nonmatching pairs (red part in Fig. \ref{fig:db:s1} and Fig. \ref{fig:db:s2}) will have significant overlap with that between matching pairs (blue part in Fig. \ref{fig:db:s1} and Fig. \ref{fig:db:s2}). But if we fuse $\mathbf{h}^{o}$ and $\mathbf{h}^{f}$ using the proposed distance boosting method in (\ref{equ:train}), $d_{\mathrm{boost}}$ between matching video pairs will tend to be much smaller than that between nonmatching video pairs, resulting that the blue histogram and red histogram in Fig. \ref{fig:db:s3} will be much less overlapping than if use only $\mathbf{h}^{o}$ or $\mathbf{h}^{f}$. This reduced overlap in turn improves detection performance, which is verified next.

\noindent \textbf{ROC comparisons against existing techniques:} Finally, we synchronize the query video using the proposed DTW-based method to get $\boldsymbol{\mathscr{V}}_q^{syn}$ as described in Algorithm \ref{alg:dtwsyn}, then apply CGO, TIRI and FH to $\boldsymbol{\mathscr{V}}_q^{syn}$. For FH, we test the performances of using $d_{\mathrm{FH}}$ and $d_{\mathrm{boost}}$, respectively. The resulting ROCs are shown in Fig. \ref{fig:roc}. FH and TIRI are competitive with FH doing slightly better. Finally, FH with distance boosting easily outperforms the alternatives.


\section{CONCLUSION}
\label{sec:conclusion}

We address the challenge of temporal desynchronization via a novel video hashing framework that involves DTW based synchronization followed by computation of a robust feature vector called flow hash (FH). Further, distance boosting is proposed to capture complementary information in FH and DTW based hash distances which delivers enhanced ROC performance under severe spatio-temporal distortions.


%

\bibliographystyle{IEEEbib}
\bibliography{refs}

\end{document}